\documentclass[aps,prb,showpacs,twocolumn,groupedaddress,floatfix]{revtex4}
\usepackage{graphicx}
\usepackage{color}
\begin{document}

\title{
Magnetic transitions induced by tunnelling electrons in individual adsorbed 
 M-Phthalocyanine molecules (M $\equiv$ Fe, Co)
}
\author{Jean-Pierre Gauyacq$^{1,2}$}
\author{Frederico D. Novaes$^{3}$}
\author{Nicol\'as Lorente$^{4}$}
 \affiliation{
$^1$ CNRS, Institut des Sciences Mol\'eculaires d'Orsay, ISMO, Unit\'e de Recherches CNRS-Universit\'e Paris-Sud,
 B\^atiment 351, Universit\'e Paris-Sud, 91405 Orsay CEDEX,
France\\
$^2$ Universit\'e Paris-Sud, Institut des Sciences Mol\'eculaires d'Orsay, ISMO, Unit\'e de Recherches CNRS-Universit\'e Paris-Sud, B\^atiment 351, Universit\'e Paris-Sud, 91405 Orsay CEDEX,
France\\
$^{3}$ Institut de Ci\`encia de Materials de Barcelona (CSIC),
Campus de la UAB, E-08193 Bellaterra, Spain\\
$^{4}$ Centre d'Investigaci\'o en Nanoci\`encia  i
Nanotecnologia (CSIC-ICN), Campus de la UAB, E-08193 Bellaterra, Spain
}
\date{\today}

\begin{abstract}
We report on a theoretical study of magnetic transitions induced by
tunnelling electrons in individual adsorbed M-Phthalocyanine (M-Pc)
molecules where M is a metal atom:
 Fe-Pc on a  Cu(110)(2$\times$1)-O surface and  Co-Pc layers
on Pb(111) islands. The magnetic transitions correspond to the change of
orientation of the  spin angular momentum of the metal ion with respect
to the surroundings and possibly  an applied magnetic field. The
adsorbed Fe-Pc system is studied with a  Density Functional Theory (DFT)
transport approach  showing that i) the magnetic structure of the Fe atom
in the adsorbed Fe-Pc is quite different from that of the free Fe atom or
of other adsorbed  Fe systems and ii) that injection of electrons (holes)
into the Fe atom in the adsorbed Fe-Pc molecule dominantly involves the Fe
$3d_{z^2}$ orbital. These results  fully specify the magnetic structure
of  the system and the process responsible for magnetic transitions. The
dynamics of the magnetic transitions induced by tunnelling electrons
is treated in  a strong-coupling  approach. The Fe-Pc treatment is
extended to the Co-Pc case. The present calculations accurately reproduce
the strength of the magnetic transitions as observed by magnetic IETS
(Inelastic Electron Tunnelling Spectroscopy) experiments; in particular,
the dominance of the inelastic current in the conduction of the adsorbed
M-Pc molecule is accounted for.
\end{abstract}

\maketitle

\section{Introduction}

The development of Scanning Tunnelling Microscopy (STM) 
recently enabled to study the case of localised magnetic excitations 
at surfaces~\cite{Heinrich,Hirjibehedin06,Hirjibehedin07,Tsukahara,XiChen,Iacovita,Fu}. 
Low-temperature STM experimental studies
revealed that, in certain cases, a local spin could be associated with
individual adsorbates at surfaces. Under the variation of the STM bias,
the tip-adsorbate junction exhibits conductance steps at well-defined
energies in the few meV energy range. These steps are attributed to the
excitation of the local spin of the adsorbate and the step energy position
yields the corresponding excitation energy, leading to magnetic IETS
(Inelastic Electron Tunnelling Spectroscopy). Revealing the existence
of nano-magnets at the atomic (molecular) level can have fascinating
consequences for the miniaturisation of electronic devices. In addition,
the possibility to determine the energy spectrum of the local spin on
the adsorbate as a function of an applied magnetic B field provides
an efficient way to quantitatively characterize the local spin and
its interaction with the underlying substrate. This makes magnetic
IETS an invaluable tool for magnetic studies of nano-objects at
surfaces. In this respect, one can mention several spectacular results
obtained in this way: evidence of spin coupling between neighbouring
atomic adsorbates (anti-ferromagnetic coupling along adsorbed Mn
chains)~\cite{Hirjibehedin06}, large magnetic anisotropy of atomic adsorbates (Mn and
Fe adsorbates on CuN)~\cite{Hirjibehedin07}, change of magnetic structure of a molecular
adsorbate on various substrates (Fe-Phthalocyanine on Cu and CuO)~\cite{Tsukahara},
evidence of super-exchange interactions~\cite{XiChen}, interaction of molecular
adsorbates with magnetic substrates~\cite{Iacovita}, {charging
of adsorbed magnetic nano-objects~\cite{Fu}}. Magnetic transitions could be
easily evidenced in these systems due to the presence of a coating on
the metal substrate that efficiently decouples the adsorbate from the
continua of metallic states. When magnetic atoms are directly adsorbed
on a metal, magnetic transitions could be observed but much broadened
by the interaction with the substrate~\cite{Balashov}. 

 Besides detailed insight into
the structure of a magnetic adsorbate on a surface, IETS experiments
bring also information on the dynamics of the excitation of a local
spin by tunnelling electrons. Indeed, when considering a conductance
spectrum as a function of the junction bias, the position of the
conductance steps yields the excited state energies and the height
of the conductance steps relative to each other and relative to the
conductance at zero bias yields the relative magnitude of the various
possible magnetic excitations. This feature is quite important for
possible future applications, since it determines how easily a local
adsorbate spin can be flipped at will by tunnelling electrons or can
be quenched by collisions with substrate electrons~\cite{FDN}. It appears that
magnetic transitions are highly 
probable~\cite{Hirjibehedin06,Hirjibehedin07,Tsukahara,XiChen}, 
much more probable than
other inelastic processes like e.g. vibrational excitation by tunnelling
electrons~\cite{Stipe,Ho,Komeda}. As an extreme example, in the case of Co-Phthalocyanine
(Co-Pc)~\cite{XiChen}, the inelastic contribution to the current is found to be three
times larger that the elastic current, whereas for vibrational excitation,
the inelastic tunnelling contribution reaches at most a few per cent~\cite{Stipe,Ho,Komeda}.

%%%%%%%%%%%%%%%%%%%
%Figure 1: 
%%%%%%%%%%%%%%%%%%%
\begin{figure}
\includegraphics[width=0.45\textwidth]{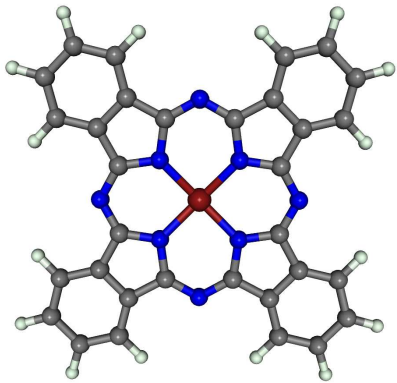}
\caption{
Scheme of an
M-Phthalocyanine molecule. In the present
study M is taken as a Fe or a Co atom, here represented in red.
The  free molecule 
has a $D_{4h}$ symmetry
which is generally reduced upon adsorption on a substrate.
The
N-atoms are depicted in blue, C atoms in grey
and H atoms in cyan. 
}
\label{figure0}
\end{figure}

A few tools have been used to describe the spectroscopy and dynamics of a
local spin on an adsorbate. First, the energy of a local spin interacting
with its environment and possibly with an applied magnetic field, B,
has been modelled efficiently with the following magnetic Hamiltonian~\cite{Yosida}:

\begin{equation}
        H = g \mu_{B} \vec{B} \cdot \vec{S}  + D S_{z}^{2} + E (S_{x}^{2}-S_{y}^{2}),
\label{hamiltonien}
\end{equation}
where $\vec{S}$  is the local spin of the
adsorbate, $g$ the Land\'e factor and $\mu_B$ the Bohr magneton.  $\vec{B}$ is an
applied magnetic field. $D$ and $E$ are two energy constants describing
the interaction of $\vec{S}$ with the substrate, i.e. the magnetic anisotropy of
the system. The Cartesian axis xyz are chosen according to the magnetic
symmetries of the system. Hamiltonian~(\ref{hamiltonien})
 describes how the adsorbate
spin, $\vec{S}$, is oriented in space due to its interaction with the substrate. The
observed magnetic excitation energies of adsorbates were very efficiently
modelled by the above Hamiltonian~\cite{Hirjibehedin06,Hirjibehedin07,Tsukahara}. 
As for the strength of the
magnetic excitations, it was first analyzed in a phenomenological way
by Hirjebehidin {\em et al}~\cite{Hirjibehedin07}. 
They showed that the relative heights of the
inelastic conductance steps were very close to the relative magnitude
of the squared matrix elements of the  operator between the initial and
final states of the transition. Nothing could be said about the relative
magnitude of the elastic and inelastic conductance. This finding was
later supported by several theoretical 
studies~\cite{Fransson,Fernandez,Persson} introducing in
the description of electron tunnelling a coupling term proportional
to the local spin. In first order perturbation theory, the inelastic
current then appears proportional to the squared matrix elements of the
coupling term between initial and final states, similarly to what was
noticed in Ref.~[\onlinecite{Hirjibehedin07}]. Recently, a completely 
different approach was introduced to
treat magnetic excitations by tunnelling electrons~\cite{Lorente}. 
It is based on the
large difference of time scale between electron tunnelling and magnetic
anisotropy of adsorbates: electron tunnelling is fast and the magnetic
anisotropy can be considered as non-active during tunnelling. One can
then describe tunnelling without the magnetic anisotropy taken into
account and simply switch the latter at the beginning and at the end of
tunnelling. In addition, DFT calculations on the two studied systems (Fe
and Mn adsorbates on CuN) showed that only one coupling scheme between
adsorbate and tunnelling electron spins is significantly contributing
to tunnelling. The magnetic excitation appears as the result of a spin
decoupling/recoupling process. This approach is non-perturbative and
can then handle the large excitation probabilities encountered in these
systems as well as make predictions for the elastic/inelastic relative
contributions to tunnelling. In addition to bringing a qualitative view
into the magnetic excitation process, it was shown to accurately account
for the strength of the magnetic excitations, in particular relative to
the elastic channel, in the case of Mn and Fe atomic adsorbates on CuN~\cite{Lorente}.
In the present work, we show how our method can be used in the
case of Fe-Pc and Co-Pc molecules (Fig.~\ref{figure0}) adsorbed
 on partly insulating
substrates. As one of the main results, this approach is shown to
precisely account for the overwhelming dominance of the inelastic
conductance over the elastic conductance observed experimentally
in these systems~\cite{Tsukahara,XiChen}.

%%%%%%%%%%%%%%%%%%%%%%%%%%%
%Figure 1
%%%%%%%%%%%%%%%%%%%%%%%%%%%
\begin{figure}
\includegraphics[angle=0,width=0.45\textwidth]{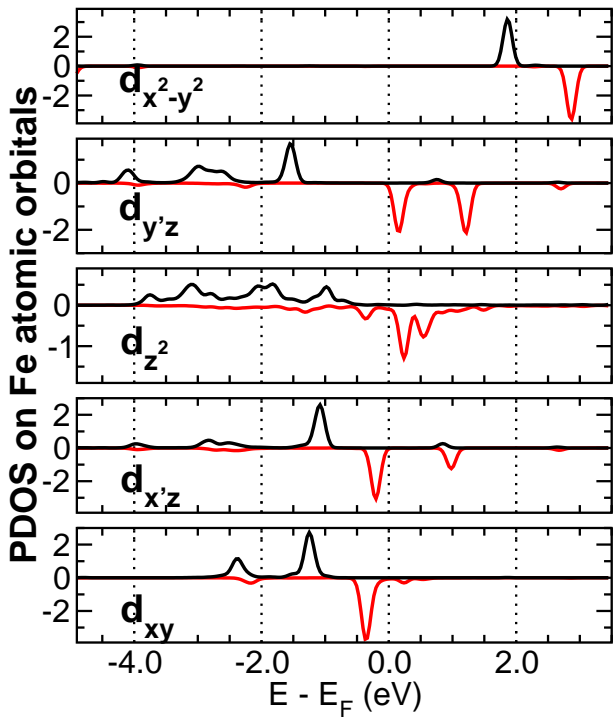}
\caption{
Projected density of states (PDOS) on the Fe $d$-atomic orbitals. For all 
the curves shown here, the positive (black) curves corresponds to the
 majority spin, and the negative (red) to the minority spin.
The $d$ orbitals are classified according to the cartesian
axes that contain the N--Cu--N axis of the molecule $(x,y,z)$, or
with respect to the surface directions: $x'$ for the $[1-10]$ and $y'$
for the $[001]$ directions. The $z$ axis is the same for both 
reference frames.
}
\label{figure1}
\end{figure}

\section{Method}
\label{2}

We consider a local spin, $\vec{S}$, localized on an adsorbate on a surface.
It is coupled to its environment by the Hamiltonian~(\ref{hamiltonien}). Diagonalizing
the Hamiltonian~(\ref{hamiltonien}) yields the various spin states of the system, their
energies, $E_n$, and the associated wave-functions, $\phi_n$. 
Hamiltonian~(\ref{hamiltonien})
is more easily written in the basis of $|S,M\rangle$ states, the eigenstates of $\vec{S}^2$
 and $S_z$, so that the anisotropy states can be written as:
\begin{equation}
\left|\phi_{n}\right\rangle = \sum_{M} C_{n,M} \left|S,M\right\rangle
\label{phi_n}
\end{equation}
In the present study, we used the $D$, $E$ and $g$ parameters in
Hamiltonian~(\ref{hamiltonien}) as determined in the analysis of the
experimental energy spectrum and from these, without any further parameter
adjustment, we derive the strength of the magnetic transitions induced
by tunnelling electrons. The various energy terms in Hamiltonian~(\ref{hamiltonien})
 are in the few meV range, so that one does not expect the corresponding interaction
to play a role during the electron tunnelling process and one can treat
this problem in the sudden approximation: treat the tunnelling without
the magnetic anisotropy and then introduce it as a frame transformation
at the beginning and at the end of tunnelling.

Electron tunnelling from the STM tip through the adsorbate
and into the substrate (in the absence of magnetic anisotropy) can be
represented by a scattering $T$ matrix, noted  $T_{Tip\rightarrow Sub}$
(and an equivalent one for the reverse tunnelling). 
It corresponds to the scattering of the
tunnelling electron by the adsorbate and it depends on the electron
energy. In the absence of magnetic anisotropy, it depends a priori on the
spin coupling between the tunnelling electron and the adsorbate, via the
exchange interaction. Thus, in the absence of significant spin-orbit interactions, it can be written in a diagonal form if we
consider $\vec{S}_T$, the total spin of the system (electron + adsorbate). Defining
$\left|S_T,M_T\right\rangle$ as
the eigenfunctions of $\vec{S}_T^2$  and $S_{T,z}$  (if $S$ is the adsorbate spin,
then $S_T = S \pm \frac{1}{2}$), we can write
formally the scattering $T_{Tip\rightarrow Sub}$  matrix 
(in the absence of magnetic anisotropy) as:
\begin{equation}
T_{Tip\rightarrow Sub}= \sum_{S_T,M_T} \left|S_T,M_T\right\rangle 
T_{Tip\rightarrow Sub}^{S_T} \left\langle S_T,M_T\right|.
\label{Tmatrix}
\end{equation}
$T_{Tip\rightarrow Sub}^{S_T}$ is a number, 
function of the electron energy. However, since, below,
we consider only the limited energy range spanned by the magnetic
excitations (up to 10 meV), the transmission probabilities $| T_{Tip\rightarrow Sub}^{S_T}|^2$ can be
considered as constant in the present study. In the sudden approximation,
the tunnelling amplitude (in the presence of magnetic anisotropy) is
written as the matrix element of the  $T_{Tip\rightarrow Sub}$ amplitude between the initial and
final states of the tunnelling process. These states are written as  
$|\frac{1}{2}, m; \phi_n \rangle $ where the
first part concerns the tunnelling electron (the electron spin is 1/2 and
$m$ is the projection of the tunnelling electron spin on the quantization
axis) and the second part concerns the local spin of the adsorbate. 
One then obtains the
amplitude, $AMP_{m,n\rightarrow m',n'}$, for a tunnelling electron induced transition from $\phi_n$ to
$\phi_{n'}$, while the tunnelling electron spin projection changes from $m$ to $m'$ as:
\begin{eqnarray}
&AMP_{m,n\rightarrow m',n'} = 
\sum_{S_T} \; T_{Tip\rightarrow Sub}^{S_T} & \nonumber \\
&\times
\sum_{M_T} \langle \frac{1}{2}, m'; \phi_n' | S_T,M_T \rangle
\langle S_T,M_T | \frac{1}{2}, m; \phi_n \rangle&
\label{amp}
\end{eqnarray}
One can see that there is a 
superposition (interference) of tunnelling amplitudes
through the different $S_T$ introduced by the magnetic anisotropy. 
Equation~(\ref{amp})
 yields the transition amplitudes in the general case, if the spin
of the tunnelling electron is registered in both the initial and final
states. Implicitly, it has been assumed above that the tunnelling electron
quantization axis is the z-axis of the adsorbate magnetic anisotropy;
situations with different quantization axis for the adsorbate and the tunnelling electron can be easily handled with an expression similar to Eq.~ (\ref{amp}).

We can now define the probability, $P$, for transitions from $\phi_n$ to $\phi_{n'}$
induced by unpolarized tunnelling electrons by summing incoherently over
the distinguishable channels:
\begin{eqnarray}
P_{n\rightarrow n'} &=& 
\frac{1}{2} \sum_{m,m'} | AMP_{m,n\rightarrow m',n'}|^2 \nonumber \\
&=& \frac{1}{2} \sum_{m,m'} | \sum_{S_T} T_{Tip\rightarrow Sub}^{S_T}   \\
&\times&
\sum_{M_T} \langle \frac{1}{2}, m'; \phi_n' | S_T,M_T \rangle
\langle S_T,M_T | \frac{1}{2}, m; \phi_n \rangle
|^2. \nonumber
\label{prob}
\end{eqnarray}
We can be a little more explicit by writing 
the total spin states,$|S_T,M_T\rangle$,
as expansions over uncoupled spin states:
\begin{equation}
|S_T,M_T\rangle = \sum_m CG_{S_T,M_T,m} |S,M=M_T-m\rangle |\frac{1}{2}, m\rangle
\label{SM}
\end{equation}
where $ |S,M\rangle$ states
correspond to the adsorbate spin states and $|\frac{1}{2}, m\rangle$
to the tunnelling electron spin. Again,  the adsorbate
and electron spins are quantized on the same axis. The CG are Clebsch-Gordan coefficients. 
Combining (\ref{SM}) and (\ref{phi_n}) we
can express the total spin states as an expansion over uncoupled products
of adsorbate magnetic anisotropy states and tunnelling electron spin:
\begin{equation}
        \left|j\right\rangle  =\left|S_{T},M_{T}\right\rangle =
        \sum_{n,m} A_{j,n,m} \left|\phi_{n}\right\rangle        \left|1/2,m\right\rangle 
\label{Ajnm}
\end{equation}

Equation~(\ref{Ajnm}) links the $ \left|j\right\rangle  =\left|S_{T},M_{T}\right\rangle$,  states appropriate to describe tunnelling without
magnetic anisotropy to the channel states of the complete tunnelling
process.  We can then rewrite the transition probability (\ref{prob}) as:
\begin{equation}
P_{n\rightarrow n'}  
= \frac{1}{2} \sum_{m,m'} | \sum_{S_T} T_{Tip\rightarrow Sub}^{S_T}   
\sum_{M_T} A_{j,n,m} A^*_{j,n',m'}|^2.
\label{P2}
\end{equation}
In the case where the $T_{Tip\rightarrow Sub}^{S_T}$
  tunnelling amplitude is dominated by a single symmetry, $S_T$
(like it was found in the case of Mn and Fe adsorbates on CuN~ \cite{Lorente},
and like it is shown below to be the case in the systems studied here),
the probability (\ref{P2}) further simplifies into:
\begin{equation}
P_{n\rightarrow n'} 
= \frac{1}{2} | T_{Tip\rightarrow Sub}^{S_T}|^2 \sum_{m,m'} |  
\sum_{M_T} A_{j,n,m} A^*_{j,n',m'}|^2.
\label{P3}
\end{equation}
This result, used in Ref.~[\onlinecite{Lorente}], is very simple,
the electronic part of
the tunnelling (the $T_{Tip\rightarrow Sub}^{S_T}$ amplitude) is factored out and the probabilities
for the different channels are simply proportional to spin-coupling
coefficients corresponding either to the magnetic anisotropy or to
the coupling between electron and adsorbate spins (the coefficients
are products of the diagonalization expansion coefficients in Eq.~ (\ref{phi_n})
 and Clebsch-Gordan coefficients).

From the transition probabilities we can write the conductance $dI/dV$ as
a function of the STM bias, $V$, as:
\begin{equation}
\frac{dI}{dV}=C_0\frac{\sum_n \Theta(V-EX_n) \sum_{m,m'}
|\sum_j A_{j,1,n} A^*_{j,n,m'}|^2}{
\sum_n \sum_{m,m'}|\sum_j  A_{j,1,n} A^*_{j,n,m'}|^2}.
\label{conductance}
\end{equation}
Expression~(\ref{conductance}) corresponds to the conductance for the 
system being initially in the ground state $n=1$. The sum over $n$ extends over all the
$| \phi_n \rangle $ states, including the ground state, 
so that the above conductance
takes
 {all contributions, elastic and inelastic,}
  into account. $EX_n$ is the excitation energy
of the magnetic level $n$, corresponding
to the eigenvalue difference of the final, $| \phi_n \rangle $,
and initial $| \phi_1 \rangle $ states. The Heavyside function, $\Theta$,
takes care of the opening of the inelastic channels at zero temperature.
$ C_0$ is the total
conductance corresponding to the  transmission amplitude 
$T_{Tip \rightarrow Sub}^{S_T}$.
It is then
a magnetism-independent conductance. Since we only consider a limited $V$
range, defined by the magnetic excitation energies, $C_0$ can be considered
as constant in the relevant $V$-range. 
$C_0$ is equal to the conductance of the
system for biases larger than all the inelastic thresholds. 
Expression~(\ref{conductance})
corresponds to the case where only one $S_T$ value actually contributes
to tunnelling so that the sum over $j$ is restricted to the corresponding
$M_T$ values. If the two $S_T$ symmetries contribute to tunnelling, a more
general expression derived from Eq.~ (\ref{P2}) has to be used.
%}

The above form (\ref{P3}) for the transition probabilities accounts for the
strength of the magnetic transitions. In fact, the transition probability
is not proportional to a coupling term as in perturbation approaches,
it appears as a simple 
{sharing} 
of the global transmission of the
tip-substrate junction, $|T_{Tip \rightarrow Sub}^{S_T}|^2$, 
among the various spin states, $|\phi_n\rangle$. 
Qualitatively,
in the magnetic excitation process, one can say that the incident
electron arrives in a given spin state, it couples with the adsorbate
spin to form a state of the total spin, $S_T$; tunnelling through the
junction occurs independently in the different total spins; at the end
of the electron-adsorbate collision, the total spin splits back into its
adsorbate and electron components, populating all the possible adsorbate
spin states. Then, the expression for the sharing process (\ref{P3}) simply
expresses angular-momentum conservation.

This kind of excitation process has been invoked in various situations
where an exchange of angular momentum is involved: resonant rotational
excitation in electron collisions on free and 
adsorbed molecules~ \cite{Abram,Teillet2000},
spin-forbidden transitions in electron-molecule 
collisions~\cite{Teillet1987} or in
atom-surface scattering~\cite{Bahrim}. 
In all cases, it leads to efficient excitation
processes. One can stress that this description of magnetic transitions is
at variance with that of the vibrational excitation induced 
by collisional electrons ~\cite{Gadzuk,Djamo} or  by tunnelling
electrons~\cite{Lorente2000,Lorente2004,Paulsson}. 
However, even in the vibrational excitation
case,  a process similar to the one discussed here for magnetic transitions does
exist. It is associated with the composition of the incident
electron linear momentum with that of the target atom. Conservation of
momentum leads to recoil of the target induced by electron scattering;
however, due to the large ratio 
between electron and nuclei masses,
the corresponding excitation is very weak and leads to very small
excitation probabilities. Vibrational excitation has then to involve
other processes such as e.g. resonant scattering, where the increase of
the collision time allows weak interactions to be 
efficient in various situations of electron-molecule collisions~\cite{Schulz,Birtwistle,Gadzuk,Djamo,Manip, Domcke,Monturet}. The
above discussion can be summarized by stressing that the angular momenta
(orbital or spin) of electrons and atoms or molecules can be of the same
order of magnitude due to quantization (a few units in many cases) so
that exchange between them is easy and has visible effects; in contrast,
because of the large electron-nuclei mass ratio, 
the electron linear
momentum is usually much smaller than that of heavy particles limiting
the efficiency of momentum exchange processes. {As a consequence,} angular
and spin degrees of freedom appear similar and the present treatment of
spin transitions is very similar to the treatment of rotational excitation
used in Ref.~[\onlinecite{Teillet2000}]
to account for the experimentally observed efficiency of
tunnelling electron in inducing molecular adsorbate rotations~\cite{Stipe2}.

%%%%%%%%%%%%%%%%%%%
%Figure 2: 
%%%%%%%%%%%%%%%%%%%
\begin{figure}
\includegraphics[width=0.45\textwidth]{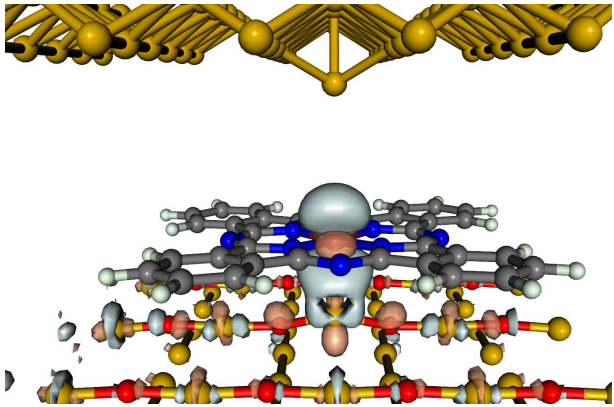}
\caption{
Amplitude of the Kohn-Sham orbital of the Fe-Pc/Cu(110)(2$\times$1)-O +
STM tip system at 0.2 eV above the Fermi level (positive in light grey
and negative in light pink). The STM tip in the upper part is made of a
Cu(110) surface   with an extra protruding atom (in gold).  The adsorbed
Fe-Pc is lying flat on the substrate (N atoms in blue, C atoms in grey
and H atoms in cyan). The substrate is in the lower part of the figure
(Cu atoms in gold and O in red). The orbital is concentrated around
the Fe atom and exhibits a strong $d_{z^2}$ character perturbed by the
interaction with the substrate.
}
\label{figure2}
\end{figure}

\section{Magnetic excitations in supported F\lowercase{e}-phthalocyanine molecules}
\label{3}

Recently Tsukahara {\em et al}~\cite{Tsukahara} 
performed a detailed magnetic IETS study of
single Fe-Phthalocyanine (Fe-Pc) molecules adsorbed on a Cu(110)(2$\times$1)-O
surface. The molecule lay flat on the surface and two adsorption
geometries were found, labelled $\alpha$ and $\beta$ 
differing by the relative
orientation of the molecule on the substrate. Clear magnetic transitions
were observed by scanning the STM tip bias and were attributed to a local
spin $S=1$ interacting with the environment and with an applied magnetic
field, $B$. The magnetic transitions were only observed when the tip was
placed above the Fe atom and were attributed to a local spin of the Fe
atom. The energies of the magnetic levels of the system as a function
of the applied field were very precisely accounted for in 
Ref.~[\onlinecite{Tsukahara}] using
Hamiltonian~(\ref{hamiltonien}). 
The parameters of the two, $\alpha$  and $\beta$ adsorption
geometries are different: $D = -3.8$ meV, $E = 1.0$ meV and $g = 2.3$ for
$\alpha$-Fe-Pc and $D = -6.9$ meV, $E = 2.1$ meV and $g = 2.4$ for 
$\beta$-Fe-Pc,
i.e. the same kind of structure but with a very different zero-field
splitting of the magnetic states. We used this modelling of the magnetic
structure to compute the strength of the magnetic excitations as
induced
by tunnelling electrons, making use of the formalism described in 
section~\ref{2}.

\subsection{Electronic structure of the Fe-phthalocyanine molecule}
\label{molecule}

A detailed description of the electronic structure of the Fe-Pc can be
found in various references (see e.g. [\onlinecite{Dale,XLu,Liao}]
 and earlier references there in). 
The $s$ outer electrons of the Fe atom are transferred to the
surrounding Pc ring leaving a central Fe$^{2+}$ ion with a $3d^6$ electronic
configuration. In the free Fe-Pc, see Fig.~\ref{figure0}, 
the Fe atom is surrounded by the
Pc ring of D$_{4h}$ symmetry, so that the Fe $d$ manifold splits into $b_{2g}$
($d_{xy}$), $e_{g}$ ($d_{zx},d_{yz}$), $a_{1g}$ ($d_{z^2}$) and 
$b_{1g}$ ($d_{x^2-y^2}$) orbitals (the $z$-axis is normal to the Pc
plane 
{and the $x$ and $y$-axis are parallel to the Fe-N interatomic axis}
). In the free Fe-Pc, the  $b_{2g}$
($d_{xy}$) orbital is fully occupied, the $b_{1g}$ ($d_{x^2-y^2}$) 
corresponds to the highest  eigenenergy because of the
large overlap with the N-atom orbitals,
 and four electrons occupy the  $e_{g}$ ($d_{zx},d_{yz}$) and
 $a_{1g}$ ($d_{z^2}$)
orbitals. Various configurations have been proposed, with small
energy differences between them (see a discussion in Ref.~[\onlinecite{Liao}]). The structure
of the other M-Pc (M= metal) is similar~\cite{Liao}, 
a change of M along the Fe,
Co, Ni, Cu and Zn sequence corresponding to the further filling of the
split $d$ manifold. 

When the Fe-Pc is adsorbed on a Cu(110)(2$\times$1)-O surface,
the D$_{4h}$ symmetry is broken. If we assume the presence of the surface to
only induce a perturbation on the structure of the free Fe-Pc, then the
$g$ subscript of the orbitals disappears and the $e_g$ orbital splits (the 
$ d_{zx}$
and $d_{yz}$ are not degenerate anymore). We will 
 use the orbital notation of the free
Fe-Pc when
 discussing the electronic
structure of adsorbed Fe-Pc, although the molecules
are distorted by the interaction with the surface.

%%%%%%%%%%%%%%%
%Figure 3
%%%%%%%%%%%%%%%
\begin{figure}
\includegraphics[angle=0,width=0.45\textwidth]{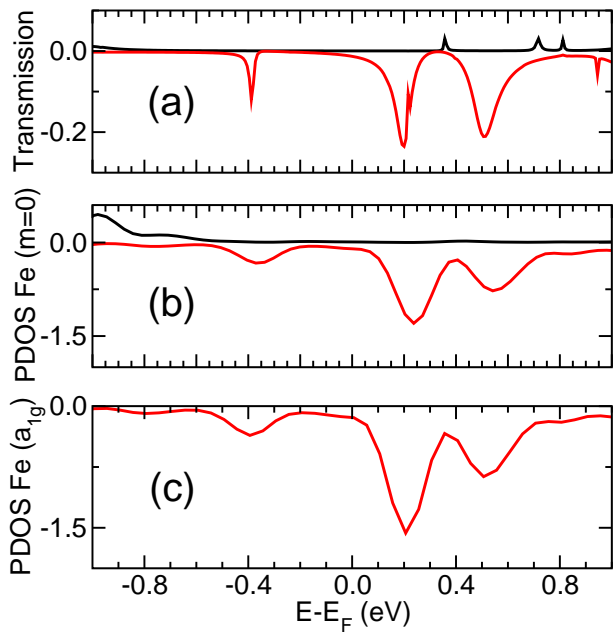}
\caption{
In panel (a), the transmission function for the system shown 
in Fig.~\ref{figure2} 
as a function of the tunnelling electron energy (referred to the Fermi energy $E_F$) for a vanishing STM bias. 
Panel (b) shows the PDOS on the Fe   $d_{z^2}$ atomic orbital.
Panel (c) shows the projection onto the $a_{1g}$ molecular orbital of the free distorted Fe-Pc
 that corresponds to the $d_{z^2}$ Fe orbital.
Positive (black) curves correspond to the majority spin, 
and the negative (red) ones to the minority spin. 
}
\label{figure3}
\end{figure}

\subsection{Density functional study of Fe-Pc adsorbed on a Cu(110)(2$\times$1)-O
surface}
\label{surface}

The ground state electronic structure configuration and the global
transmission, $T_{Tip \rightarrow Sub}^{S_T}$ 
for the tunnelling from an STM tip { to the substrate} passing through
the Fe-Pc molecule were obtained by density functional theory (DFT) 
simulations, performed with
the {\sc Siesta} and {\sc Transiesta} codes~\cite{Siesta,Transiesta}.  
We have used the generalized gradient approximation~\cite{PBE} 
for the exchange-correlation functional. We only studied the $\beta$ adsorption geometry
of the Fe-Pc on Cu(110)(2$\times$1)-O, which has a higher symmetry than the $\alpha$ geometry. 
The electronic structure in the  $\alpha$ and $\beta$  geometries are assumed to be equivalent.

Starting with the electronic structure analysis for the adsorbed molecule,
{\sc Siesta} uses atomic orbitals for the basis set, and the projection of the
density of states onto the Fe atomic orbitals 
%(we have used an optimized DZP basis for the Fe-Pc)  
reveals {an open shell structure with two unpaired electrons, i.e.} a $S=1$ configuration 
($d_{xy}^2 d_{y'z}^2 d_{z^2}^1 d_{x'z}^1$), as shown
in Fig.~ \ref{figure1}. 
  Note that the $\beta$ adsorption geometry corresponds to the Fe-N axis rotated by 45$^\circ$
 from the symmetry axis of the Cu(110) surface. As a consequence, the splitting of the $e_g$ orbital
 by the substrate field involves $x'$ and $y'$-axis rotated by 45$^\circ$ from the $x$ and $y$-axis that 
correspond to the splitting of the $d$ manifold by the interaction with the Pc ring. The projections in Fig.~\ref{figure1}
 were then performed on the appropriate symmetry orbitals i.e. the $d_{xy}$, $d_{z^2}$, $d_{x^2-y^2}$, $d_{zx'}$ and $d_{y'z}$ orbitals.
More quantitatively, 
the computed spin polarization is  1.85 unpaired electrons. The corresponding
 spin  is 
roughly $S=1$, in good agreement with the experiment~\cite{Tsukahara}. 
Due to its large overlap with the surface electronic structure,
the $d_{z^2}$ orbital is the one that hybridizes the most with
the substrate.

%%%%%%%%%%%%%%%
%Figure 4
%%%%%%%%%%%%%%%
\begin{figure}
\includegraphics[width=0.45\textwidth]{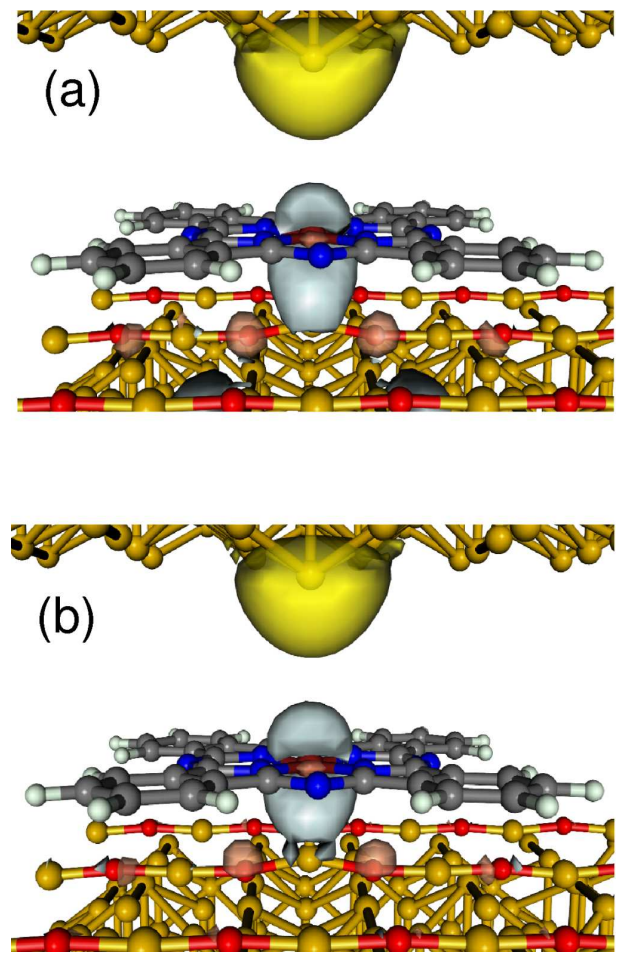}
\caption{
Transmission eigenchannel corresponding to the
largest transmission amplitude for 
(a) $E=E_F$ where $E_F$ is the system's Fermi energy and 
(b) $E=E_F + 0.2$ eV. Light grey (light pink) color corresponds to the positive (negative)
 imaginary part of the eigenchannel amplitude coming from the STM tip.
 In gold color, the positive real part. Note that the isosurfaces were chosen different in 
(a) and (b) because of the large difference in transmission probability.
 The transmission channel exhibits   a strong $a_{1g}$ ($d_{z^2}$) character around the Fe atom.
}
\label{figure4}
\end{figure}

Figure~\ref{figure2}
 shows the real space plot of the  Kohn-Sham orbital at 0.2 eV above the Fermi level for the
molecule+surface+STM tip system. This orbital corresponds to the peak in the PDOS onto the $d_{z^2}$  
Fe orbital (see Fig.~\ref{figure1}). The strong  $d_{z^2}$ character of the orbital seen in 
Fig.~\ref{figure2} confirms that this orbital corresponds to the $a_{1g}$ orbital perturbed by the substrate. 
The intuitive notion that, by reaching further out along the $z$-axis, this orbital would 
contribute more to the junction transmission is
supported by comparing the transmission function 
 (panel (a)
of Fig.~\ref{figure3})
with the
PDOS onto the $d_{z^2}$ Fe atomic orbital  (b), and PDOS onto the molecular
orbital $a_{1g}$ of the Fe-Pc (c). 
Note that, 
for this projection, we used the $a_{1g}$ orbital of the free Fe-Pc molecule, 
with the distorted atomic configuration of the adsorbed molecule.
From this, it is clear that, at the Fermi level, the transmission is
essentially made of the tail of the $a_{1g}$ resonance due to the $d_{z^2}$ Fe  orbital,
and that only the minority spin contributes to the tunneling tranmission. 
In the transmission calculations, the tip-molecule distance was set at a small distance, 5~\AA, with a large enough transmission probability to visualize  the electron transmission eigenchannel easily.

This conclusion is further confirmed
by computing the transmission eigenchannels. For that, we have used
the {\sc Inelastica} code~ \cite{Inelastica}. The  $d_{z^2}$  orbital is the largest contributor to S-matrix
eigenchannel {dominating} the 
transmission of electrons from the STM tip (here
represented by an atom on a Cu(110) semi-infinite electrode) 
placed above the
molecule.
As can be seen in Fig.~\ref{figure4}, the character of the
{ dominant transmission} eigenchannel is the same at the Fermi energy, $E_F$, Fig.~\ref{figure4}(a)
 and at $E=E_F + 0.2$ eV, Fig.~\ref{figure4} (b),
the peak of the transmission resonance. These plots nicely correspond to the
density plot of the Kohn-Sham eigenstate, Fig.~\ref{figure2}.

%%%%%%%%%%%%%%%%%%%%%%%%%%%%%%%%%
% Fig 5
%%%%%%%%%%%%%%%%%%%%%%%%%%%%%%%%%
\begin{figure}
\includegraphics[width=0.45\textwidth]{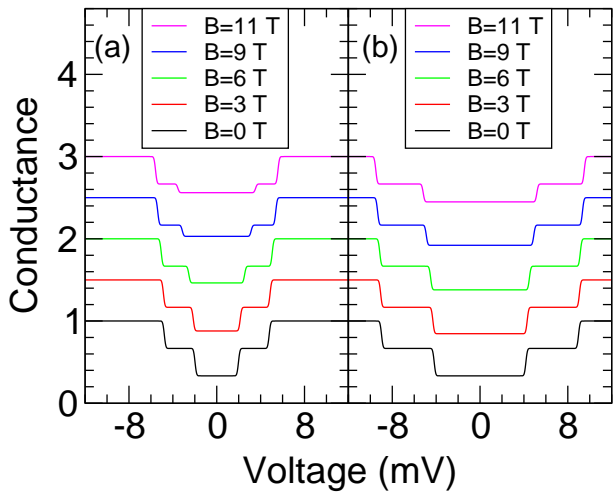}
\caption{
Computed conductance  for the (a) $\alpha$ and (b) $\beta$ configuration
of an adsorbed Fe-Pc molecule on  a Cu(110)(2$\times$1)-O
surface.
The conductance has been normalized to 1 at large bias
and the curves for the various $B$ fields have been offset 
by 0.5 for representation purposes.
The  B field is oriented  vertical to the sample with
magnitude 0, 3, 6, 9 and 11 T as indicated in the graph.
\label{figure5}}
\end{figure}

\subsection{Orbital picture of the electron tunnelling process} 

From the above DFT study, we can conclude that  the Fe-Pc on a
Cu(110)(2$\times$1)-O surface, 
is associated to the triplet electronic configuration 
$d_{xy}^2 d_{y'z}^2 d_{z^2}^1 d_{x'z}^1$  
consistently with experiment~ \cite{Tsukahara}. The
interaction between the $d$ manifold and the surroundings is seen
to split completely the $d$ manifold, leading to five non-degenerate
orbitals; the orbital angular momentum is then completely quenched,
in the absence of significant spin-orbit couplings~\cite{Yosida}
 and this justifies
the discussion, used in Ref.~\cite{Tsukahara} and here, 
of the magnetic anisotropy in
terms of the spin angular momentum orientation. One can also notice that
the Fe spin is also partially quenched from its free atom value ($S =
2$), again due to the interaction with the surroundings that induces a
large upward energy shift of the $d_{x^2-y^2}$ orbital. 
The present DFT study also
shows that when an STM tip is placed above the Fe atom, tunnelling
dominantly involves the $d_{z^2}$  ($a_{1g}$) orbital. So when an electron is sent
from the tip on the Fe, it involves the  {$d_{xy}^2 d_{y'z}^2 d_{z^2}^2 d_{x'z}^1$}
 {transient} configuration, i.e. the total
spin of the electron-adsorbate scattering intermediate is $S_T = 1/2$.
Similarly, if a hole 
is sent from the tip to the Fe
atom, tunnelling involves the   {$d_{xy}^2 d_{y'z}^2 d_{z^2}^0 d_{x'z}^1$} {transient}
configuration, i.e. the total spin of the
electron-adsorbate scattering intermediate is again  $S_T = 1/2$. Thus, of
the two possible symmetries for the electron-adsorbate scattering ($S_T =
1/2$ or $3/2$), 
$S_T = 1/2$ is the prevailing
symmetry in the tunnelling process.

{
In addition, transmission through the  $d_{z^2}$ orbital will dominate 
in a constant-current STM image and 
generate a bright spot at the Fe centre}. This is consistent with the
observation in Ref.~[\onlinecite{Tsukahara}] 
of the Fe-Pc molecule as a bright spot at the centre
of a clover leaf. The bright spot corresponds to
 the  $d_{z^2}$  Fe orbital and the clover leaf is given by
the contribution from other orbitals localized on the Pc ring (see in~[\onlinecite{XLu}]
a discussion of the link between bright M atoms in M-Pc STM images and their
$d$ orbitals). Furthermore, the magnetic transitions in this system were
also found~\cite{Tsukahara} to be localized in the same region  
about the Fe atom, {which we attribute to the $d_{z^2}$ orbital.}

\subsection{Magnetic excitation processes}	

The inelastic conductance of the Fe-Pc has been computed using 
expression~(\ref{conductance}) for the $\alpha$ and $\beta$
 adsorption geometries, using the spin
parameters determined in the DFT study. The conductance is shown
in Fig.~\ref{figure5}(a) for $\alpha$ configuration
 and Fig.~\ref{figure5} (b) for the $\beta$ configuration, as a function of the STM bias 
for various values of the applied
magnetic field, $B$, along the $z$-axis. The conductance has been {normalized to 1} at large bias
and the curves for the various $B$ fields have been {offset 
by 0.5}. In the calculation, a Gaussian broadening of 0.25 meV
was introduced to mimic various broadening effects. The conductance
spectra resemble very much those measured by Tsukahara 
{\em et al}~\cite{Tsukahara}, with
well-marked steps at the magnetic excitation thresholds (two inelastic
thresholds for this $S=1$ system because anisotropy splits
the triplet state in three states~\cite{Tsukahara}) 
and very significant contributions
from the inelastic currents. Indeed, the conductance is dominated by
inelastic tunnelling at large bias! 

The differences between the $\alpha$  and $\beta$
 adsorption geometries are also well reproduced. One can also stress
that, in the present approach, the magnetic excitations do not modify the
maximum value of the global conductance $C_0$, 
which is then independent of changes
in the magnetic structure and in particular independent of the applied $B$
field. This appears clearly in Figs.~\ref{figure5} (a) and (b) %\ref{figure6}
 as well as in Fig.2a and 2b of Ref.~[\onlinecite{Tsukahara}],
confirming our view of the magnetic excitation process as a sharing of
the global conductance over the various possible magnetic channels.

\begin{figure}
\includegraphics[width=0.45\textwidth]{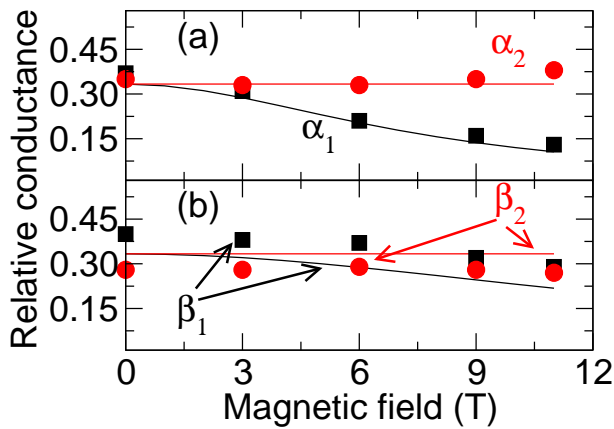}
\caption{
Relative inelastic step heights in the conductance
as a function of the magnetic field B, for (a) $\alpha$  and
(b) $\beta$ configurations. $\alpha_1$ and $\alpha_2$
refer to the first and second excitation steps  for the $\alpha$ configuration
respectively. Analogously, $\beta_1$ and $\beta_2$ refer
to the first and second excitation steps for the $\beta$ configuration.
The experimental data points are represented
with black squares for the first excitation and
as red circles for the second one and are taken from the supplemental
material of Ref.~[\onlinecite{Tsukahara}].
\label{figure7}}
\end{figure}

The relative magnitude of the elastic and inelastic channels in this
system are further illustrated in Fig.~\ref{figure7}
 which presents the magnitude
of the inelastic conductance steps for 
%a total conductance at large bias equal to unity. 
$C_0=1$.
The two inelastic step heights are $\alpha_1$ and $\alpha_2$ ($\alpha_1$
for the lowest threshold) for the $\alpha$  geometry (and similarly for 
the $\beta$ one).
The elastic conductance is then equal to the global conductance
minus the inelastic ones, (1 - $\alpha_1$ - $\alpha_2$)
with this definition. 

The present theoretical results as functions of
the applied $B$ field are compared with the data of 
Tsukahara {\em et al}~\cite{Tsukahara}. As
a first remark, inelastic tunnelling contributes significantly to the
tunnelling current: typically for $B = 0$ T, the inelastic current is
equal to twice the elastic current. Second, we can see that the present
theoretical results reproduce very well the {relative magnitude of the three contributions to the 
conductance (elastic and two inelastic).}
%, with respect to the each other and with respect to the
%elastic one}
In particular, the variation with $B$ is well accounted for. 

The variation of the strength of the magnetic excitation with $B$ is in fact
reflecting the change of the magnetic structure of the adsorbate. For $B
= 0$, the structure of the conductance curve
 is due to the anisotropy imposed by the substrate to
the Fe-Pc molecule; although the level positions are different in the
$\alpha$  and $\beta$
adsorption geometries, the eigenstates of Hamiltonian (\ref{hamiltonien}) 
at $B = 0$ are
the same in both geometries (see Tsukahara {\em et al}~\cite{Tsukahara}) 
and consequently,
the excitation probabilities are the same, as seen in Fig.~\ref{figure7}. 
The effect of a finite $B$ field is to decouple the Fe-Pc spin 
from the substrate and
to tend to a Zeeman limit at large $B$. This decoupling is easier for the
$\alpha$-geometry, because of a weaker transversal
anisotropy $E$ (given above). In the limit of  very
large $B$ (not fully reached here), the $\phi_n$ states reduce to
$|S,M\rangle$  states with
the ground state corresponding to $M = -1$. The only allowed transition is
to the ($M = 0$) excited state, so that only one inelastic step remains
in the conductance spectrum. Its height is given by 
 the modulus square of a Clebsch-Gordan
coefficient and in this limit, the inelastic current is equal to one
third of the total current.

%%%%%%%%%%%%%%%%%%%%%%%%%%%%%%%%%
% Fig 6
%%%%%%%%%%%%%%%%%%%%%%%%%%%%%%%%%

\section{Magnetic transitions in supported C\lowercase{o}-Phthalocyanine molecules}
\label{CoPc}

{
Superposed layers of Co-Pc molecules on Pb(111) islands were studied
experimentally by magnetic IETS and revealed the existence of
superexchange interactions~\cite{XiChen}}. Magnetic
excitation of the Co-Pc molecules 
were observed but only for several adsorbed layers
of Co-Pc molecules. No excitation was observed in the single layer case,
suggesting that
either the spin of the Co-Pc molecular layer
 lying directly on the Pb surface
was quenched by the interaction with the substrate or the possible
 magnetic
excitations were very short lived~\cite{XiChen}. 
The Co-Pc spin was found to be $S = 1/2$, 
and when several layers are stacked one on top of each other,
with a stacking angle of 60$^\circ$, the spins of the Co-Pc molecule
 in the second and
outer layers couple together in a way well-described by a Heisenberg
Hamiltonian, with an anti-ferromagnetic exchange coupling, $J$, of the
order of 18 meV~\cite{XiChen}.

Very strong magnetic excitations were reported in the
multi-layer case~\cite{XiChen}.  The
present formalism can be used to predict the magnitude of the magnetic
excitation in such systems. If we neglect the magnetic anisotropy of
adsorbed Co-Pc, the only ingredients in our approach are the adsorbate
spin $S$, the intermediate total spin $S_T$, and the anti-ferromagnetic
interaction $J$ for the stacked molecules. Both $S$ and $J$ are
known from experiment~\cite{XiChen}. 
We did not perform a detailed DFT study as for
the Fe-Pc case. 
We assumed that we can {extrapolate the electronic structure from Fe-Pc to Co-Pc~\cite{Liao}}, 
in both cases the Pc molecules being partly decoupled from the underlying metal.
As discussed in Ref.~\onlinecite{Liao}, 
an M-Pc series is formed by the various incomplete
$d$-shell metals and the Co structure corresponds to adding a $d$ electron
to the Fe case. With the open shell structure of Fe-Pc outlined above
{($d_{xy}^2 d_{y'z}^2 d_{z^2}^1 d_{x'z}^1$)}, 
this leads to a doublet configuration of the Co-Pc electronic structure, 
independently of the orbital
in which the electron is added. This is perfectly consistent with the
experimental observation. Then, we can assume that tunnelling involves
minority spin electrons because
the molecular electronic structure at the Fermi level is given by
the last partially-occupied $d$ orbital also in this case. Hence, the
intermediate spin state is $S_T = 0$, both for electrons and holes.

In the case of a single active Co-Pc molecule (two adsorbed layers),
an applied magnetic field is needed to perform a magnetic IETS
experiment, because of the molecular spin $S = 1/2$, 
see Ref.~[\onlinecite{XiChen}]. For a finite $B$
field, a Zeeman splitting corresponding to a Land\'e factor 1.88 has
been observed~ \cite{XiChen}. 
Assuming a simple Zeeman structure, i.e. no magnetic
anisotropy induced by the substrate, the above formalism, at finite $B$,
predicts a single inelastic peak with an inelastic current equal to the
elastic current (note that this result is independent of the value of the
Land\'e factor). This is consistent with the observations~\cite{XiChen}, 
which reported
an increase of the current around 110-120 \% at the inelastic threshold
(see Fig.~2c in Ref.~[\onlinecite{XiChen}]).

In the case of three molecular layers, two stacked molecules in the
upper two layers interact via an anti-ferromagnetic coupling, so that the
ground state of the two-molecule system is a spin 0 state while the spin
1 states are excited states (we neglect a possible anisotropy induced by
the substrate). Magnetic excitations of the system by tunnelling electrons
for a vanishing $B$ have been observed experimentally as a sharp step in the
conductance~\cite{XiChen}. 
The above formalism predicts at $B = 0$, in the absence of any
anisotropy induced by the substrate, a single conductance step at finite
bias (the transition from $S = 0$ to $S = 1$), the inelastic current being
three times larger than the elastic one. An accurate quantitative comparison
with experiment of the step height is difficult because of the non-flat
behaviour of the global conductance in this system (see Ref.~[\onlinecite{XiChen}]); nevertheless, our
prediction of  a 300 \% increase of the conductance at the inelastic threshold is
in good agreement with the experimental data. For finite $B$, the $S=1$
excited states split into a Zeeman structure and the present formalism
predicts three steps in the conductance, each of them with a height
equal to the elastic conductance. Again, this Zeeman limit corresponds
to what is observed experimentally for large $B$ (see e.g. Fig.~3b in 
Ref.~[\onlinecite{XiChen}]). 
As the main feature of this system, one can stress the
extremely large contribution of the inelastic current compared to that
of the elastic current in the total current: 
the present theory predicts a total inelastic
current three times larger than the elastic one, in good agreement
with what is observed experimentally. One can also emphasize that in the
Co-Pc case, we only discuss the Zeeman structure limit, so that the only
ingredients in our approach are the value of the local spin ($S = 1/2$)
and of the spin of the electron+adsorbate system ($S_T = 0$) and all the
predictions of relative magnitudes of elastic and inelastic currents
are directly obtained from Clebsch-Gordan coefficients.

\section{Summary and concluding remarks}

The orientation of the magnetic moment of individual adsorbates on a
surface leads to a magnetic structure with excitation energies in the
few meV range. Recent experimental developments in the low-Temperature
 Inelastic
Electron Tunnelling Spectroscopy allowed the direct observation of
transitions among these magnetic states in the case of magnetic atoms
adsorbed on a metal with a decoupling coating in between. The present
paper reports a theoretical study of magnetic excitations induced
by tunnelling electrons in metal-phthalocyanine molecules adsorbed on
surfaces (Fe-Pc on Cu(110)(2$\times$1)-O and Co-Pc stacked on Co-Pc on Pb)
that have been the subject of recent experimental studies~\cite{Tsukahara,XiChen}. It is
based on a theoretical framework recently introduced~\cite{Lorente}
 to treat tunnelling
electron-induced magnetic transitions and {on a DFT calculation to characterize the electronic structure of the Fe-Pc molecule on Cu(110)(2$\times$1)-O system.}

%needs the evaluation of the electronic-structure
%is associated with ab initio DFT calculations for 
%of the Fe-Pc molecule on Cu(110)(2$\times$1)-O; here,
% we use DFT-based calculations to characterize this system.
%}

Our approach determines the strength of the magnetic transitions
induced by tunnelling electrons when the STM tip is placed on top
of the magnetic atom. The input ingredients in our calculations are
the magnetic Hamiltonian describing the interaction of  the adsorbate
magnetic moment with its environment and the total spin of the tunnelling
electron-adsorbate system.  The magnetic Hamiltonian has been taken from
its parameterisation using the experimental results on the energy spectrum
of magnetic levels~\cite{Tsukahara,XiChen}. 
In the Fe-Pc on Cu(110)(2$\times$1)-O case, a DFT study
determined the electronic structure of the adsorbed Fe-Pc molecule 
together with
the symmetry and spin structure of the tunnelling electron. An excellent
account of the experimental findings was obtained; in particular, the
extremely large magnetic excitation probabilities (inelastic contribution
dominating over the elastic one in the tunnelling current) were confirmed.

The Fe atom in the adsorbed Fe-Pc has  electronic and magnetic structures quite different from those of a free Fe atom and  of Fe adsorbed on CuN studied earlier~\cite{Hirjibehedin07,Lorente}. The interaction of the $d$-manifold with the Pc ring and with the CuO substrate results in a full splitting of the manifold and in a  spin state different from the atomic case. Then a Fe atom inside an adsorbed Fe-Pc molecule~\cite{Tsukahara} appears very differently in a  magnetic IETS experiment compared to  the case of an adsorbed Fe atom~\cite{Hirjibehedin07}.

The magnetic transitions appear to be much more probable than other
inelastic processes studied earlier, such as vibrational excitation
of the adsorbate~\cite{Stipe,Ho,Komeda}. 
In the most spectacular case (Co-Pc, see above and
Ref.~[\onlinecite{XiChen}]), 
the inelastic contribution to the tunnelling current is three
times larger than the elastic one. The present approach explains this
striking difference. Since electron tunnelling occurs on a short time
scale compared to magnetic anisotropy, one can treat electron tunnelling
independently of the magnetic anisotropy in the sudden approximation. The
magnetic transitions then appear as the result of a change of coupling
scheme for the adsorbate spin: coupling to the adsorbate environment in
the initial and final states and coupling to the tunnelling electron
spin via exchange interactions for the tunnelling process. The strength
of the magnetic transitions is then determined by spin coupling
coefficients (such as e.g. Clebsch Gordan coefficients) and the present
approach reduces to computing how a magnetism-independent tunnelling
current is shared among the various magnetic states, i.e. how a total
magnetism-independent current is shared between elastic and inelastic
parts. The importance of a given magnetic transition is then linked to
the weight of the initial and final states in the magnetism-independent
collision intermediate and it can thus be very large. In particular, it
does not depend on the strength of an interaction coupling initial and
final states during electron tunnelling. Our approach is, thus, perfectly
well-adapted to treat situations like the present ones, where tunnelling
appears to be dominated by inelastic effects.

The strength of the magnetic transitions thus appears to be the direct
consequence of the spin coupling scheme of the system. The variation
of the magnetic transitions with an applied magnetic field, $B$, follows
the variation of the adsorbate magnetic structure with $B$, basically the
switch from a magnetic anisotropy induced by the adsorbate environment
to a Zeeman structure, i.e. the decoupling of the adsorbate spin from
its environment by the $B$-field action (see e.g. Fig.~\ref{figure5}). 
In this way,
the present study of the magnetic transitions strength as a function of $B$ 
further strengthens the knowledge of the adsorbate magnetic structure as
it can be derived from the analysis of the experimental energy spectrum
of the magnetic states. Basically, the analyses of the strength of the
magnetic transitions and of the energy spectrum are probing the same
properties of the system.

\begin{acknowledgments}
We are grateful to Prof. Noriaki Takagi and Prof. Maki Kawai for 
extended discussions on the experimental data. We thank Prof. Magnus
Paulsson for his help in some of the calculations.
Financial support from the spanish MICINN through grant FIS2009-12721-C04-01
is gratefully ackowledged. N.L. thanks Universit\'e Paris-Sud for an
Invited Professorship and for its hospitality. F.D.N. would like to thank the 
Centro de Supercomputaci\'on de Galicia (CESGA) for providing computational resources.
\end{acknowledgments}

%\include{biblios}
%Références


\begin{thebibliography}{99}
\bibitem{Heinrich} A. J. Heinrich, J. A. Gupta, C. P. Lutz and D. M. Eigler,
Science {\bf 306}, 466 (2004).


\bibitem{Hirjibehedin06} C. F. Hirjibehedin, C. P. Lutz and A. J. Heinrich,
Science {\bf 312}, 1021 (2006).

\bibitem{Hirjibehedin07} C. F. Hirjibehedin, C.-Y. Lin, A. F. Otte, M. Ternes,
C. P. Lutz, B. A. Jones and A. J. Heinrich,
Science {\bf 317}, 1199 (2007).

\bibitem{Tsukahara} N.~Tsukahara, K.~Noto, M.~Ohara, S.~Shiraki, 
N.~Takagi, Y.~Takata, J.~Miyawaki, M.~Taguchi, A.~Chainani, 
S.~Shin and M.~Kawai Phys. Rev. Lett. {\bf 102}, 167203 (2009).

\bibitem{XiChen} Xi Chen, Y.-S.~Fu, S.-H.~Ji, T.~Zhang, P.~Cheng, 
X.-C.~Ma, X.-L.~Zou, W.-H.~Duan, J.-F.~Jia and Q.-K.~Xue, 
Phys. Rev. Lett. {\bf 101}, 197208 (2008).  

\bibitem{Iacovita} C.~Iacovita, M.~V.~Rastei, B.~W.~Heinrich, 
T.~Brumme, J.~Kortus, L.~Limot and J.~P.~Bucher,  Phys. Rev. Lett. {\bf 101},  116602 (2008).

\bibitem{Fu} Y.-S. Fu, T. Zhang, S.-H. Ji, X. Chen, X.-C. Ma, J.-F. Jia and Q.-K. Xue, Phys. Rev. Lett. {\bf 103},  257202 (2009).

\bibitem{Balashov} T.~Balashov, T.~Schuh, A.~F.~Takacs, A.~Ernst,
 S.~Ostanin, J.~Henk, I.~Mertig, P.~Bruno, T.~Miyamachi, S.~Suga
 and W.~Wulfhekel,
 Phys. Rev. Lett.  {\bf 102}, 257203 (2009).

\bibitem{FDN}
F.~D.~Novaes, N.~Lorente and J.-P.~Gauyacq, {\em to be published.}


\bibitem{Stipe} B. C. Stipe, M. A. Rezai, and W. Ho,
Science {\bf 280}, 1732 (1998).


\bibitem{Ho} W. Ho, J. Chem. Phys. {\bf 117}, 11033 (2002).

\bibitem{Komeda} T. Komeda, Progress in Surf. Sci. {\bf 78}, 41 (2005)
.

\bibitem{Yosida}
K.Yosida {\em Theory of magnetism}, Springer series in solid-state science (Springer, Berlin, Heidelberg, 1996)

\bibitem{Fransson}  J. Fransson, Nano Lett. {\bf 9}, 2414 (2009).

\bibitem{Fernandez} J. Fern\'andez-Rossier,
Phys. Rev. Lett. {\bf 102}, 256802 (2009).


\bibitem{Persson} M. Persson, 
Phys. Rev. Lett. {\bf 103}, 050801 (2009).


\bibitem{Lorente} N. Lorente and J.-P. Gauyacq,
Phys. Rev. Lett. {\bf 103}, 176601 (2009).

\bibitem{Abram}
R.A.Abram and A.Herzenberg, Chem. Phys. Lett.
{\bf 3}, 187 (1969).

\bibitem{Teillet2000} D.~Teillet-Billy, J.-P.~Gauyacq and M.~Persson, Phys.Rev. B {\bf 62},  R 13306 (2000).

\bibitem{Teillet1987} D.~Teillet-Billy, L.Malegat and J.-P.~Gauyacq,
J. Phys. B {\bf 20}, 3201 (1987).

\bibitem{Bahrim} B. Bahrim, D.~Teillet-Billy and J.-P.~Gauyacq,
Phys. Rev. B {\bf 50}, 7860 (1994).

\bibitem{Gadzuk} J.W. Gadzuk, J. Chem. Phys. {\bf 79}, 3982 (1983).

\bibitem{Djamo} V. Djamo, D.Teillet-Billy and J.P.Gauyacq, Phys. Rev. Lett. {\bf 71}, 3264, (1993). 

\bibitem{Lorente2000} N. Lorente and M. Persson, Phys. Rev. Lett.
{\bf 85}, 2997 (2000).

\bibitem{Lorente2004} N. Lorente, App. Phys. A {\bf 78}, 799 (2004).

\bibitem{Paulsson} M. Paulsson, T. Frederiksen, H. Ueba,
N. Lorente and M. Brandbyge, Phys. Rev. Lett. , 226604 (2008).

\bibitem{Schulz} G. J. Schulz, Rev. Mod. Phys. {\bf 45}, 423 (1973).

\bibitem{Birtwistle} D.J. Birtwistle and A. Herzenberg, J. Phys. B
{\bf 4}, 53 (1971).

\bibitem{Manip} A. J. Mayne, F. Rose and G. Dujardin, Faraday
Discuss. {\bf 117}, 241 (2000).

\bibitem{Domcke} M. Cizek, M. Thoss, and W. Domcke, Phys. Rev. B
{\bf 70}, 125406 (2004).

\bibitem{Monturet}  S. Monturet and N. Lorente,
 Phys. Rev. B {\bf 78}, 035445 (2008).



\bibitem{Stipe2} B.~C.~Stipe, M.~A.~Razaei and W.~Ho, Science {\bf 279}, 1907 (1998).

\bibitem{Dale} B.~W.~Dale, R.~J.~P.~Williams, C.~E.~Johnson
 and T.~L.~Thorp, J. Chem. Phys. {\bf 49}, 3441 (1968).

\bibitem{XLu}  X.~Lu and K.~W.~Hipps, J. Phys. Chem. B {\bf 101},
5391 (1997).

\bibitem{Liao} M.~S.~Liao and S.~Scheiner, J.~Chem.~Phys. {\bf 114},
 9780 (2001).

\bibitem{Siesta} J.~M.~Soler, E.~Artacho, J.~D.~Gale, A.~Garcia, J.~Junquera, P.~Ordejon, D.~Sanchez-Portal,
         J.~Phys.:~Condens.~Matter {\bf 14}, 2745 (2002).

\bibitem{Transiesta}  M. Brandbyge, J. L. Mozos, P. Ordejon, J. Taylor and K. Stokbro,
Phys. Rev. B {\bf 65}, 165401 (2002); F. D. Novaes, A. J. R.
da Silva and A. Fazzio, Brazilian Journal of Physics, {\bf 36}, 799
(2006). Starting from version 3.0b, {\sc Siesta} includes the {\sc Transiesta} module, see http://www.icmab.es/siesta/

\bibitem{PBE} J. Perdew, K. Burke, and M. Ernzerhof, Phys. Rev. Lett. {\bf 77}, 3865 (1996). 
\bibitem{Inelastica}     T. Frederiksen, M. Paulsson, 
M. Brandbyge and A.-P. Jauho, Phys. Rev. B {\bf 75}, 205413 (2007);
    M. Paulsson and M. Brandbyge, Phys. Rev. B {\bf 76}, 115117 (2007). The source code can be downloaded from http://sourceforge.net/projects/inelastica 

\end{thebibliography}
\end{document}